\documentclass[12pt,preprint2]{aastex}
\usepackage{emulateapj5,apjfonts,epsfig,onecolfloat5}

\slugcomment{Submitted August 8, 2002; Accepted Oct. 3, 2002; ApJL, in press}

\shorttitle{Proton Cyclotron Features in SGR~1806-20}
\shortauthors{Ibrahim, Swank \& Parke}

\begin{document}
\twocolumn[

\title{New Evidence for Proton Cyclotron Resonance in a Magnetar Strength Field from SGR~1806-20}

\author{Alaa I. Ibrahim,$^{1,2}$ Jean H. Swank$^{1}$ \& William Parke$^{2}$} 
\affil{$^{1}$ NASA Goddard Space Flight Center, Laboratory for High Energy Astrophysics, 
Greenbelt, MD 20771 \\ 
$^{2}$ Department of Physics, The George Washington University, Washington, D.C. 20006\\
Alaa@milkyway.gsfc.nasa.gov}

\begin{abstract}

A great deal of evidence has recently been gathered in favor of the picture that Soft 
Gamma Repeaters and Anomalous X-Ray Pulsars are powered by ultra-strong magnetic fields
($B>10^{14}$ G; i.e. {\it magnetars}). Nevertheless, present determination of the magnetic 
field in such magnetar candidates has been indirect and model dependent. A key prediction
concerning magnetars is the detection of ion cyclotron resonance features, which would offer a 
decisive diagnostic of the field strength. Here we present the detection of a 5 keV absorption 
feature in a variety of bursts from the Soft Gamma Repeater SGR~1806-20, confirming our initial 
discovery \citep*{ib:2002} and establishing the presence of the feature in the source's burst 
spectra. The line feature is well explained as proton cyclotron resonance in an ultra-strong 
magnetic field, offering a direct measurement of SGR~1806-20's magnetic field ($B\approx10^{15}$~G) 
and a clear evidence of a magnetar. Together with the source's spin-down rate, the feature also 
provides the first measurement of the gravitational redshift, mass and radius of a magnetar.

\end{abstract}

\keywords{Pulsar: Individual (SGR~1806--20) --- Stars: Magnetic Fields --- Stars:
Neutron --- Stars: Magnetar --- X-Rays: Bursts --- Gamma Rays: Bursts }

]

\section{Introduction}

Soft Gamma Repeaters (SGRs) are a rare, enigmatic type of pulsars that are distinguished by 
the hallmark of episodic emission of short ($\sim 0.1$~s), hyper-luminous ($\sim 
10^{39}-10^{42}$~erg$\,{\rm s}^{-1}$) bursts of soft $\gamma$-rays (see e.g. \citealt{hu:2000} for 
reviews). Less frequently, SGRs also emit giant flares that last for hundreds of seconds, exhibit 
remarkable pulsations, and even have measurable effects on the earth's outer atmosphere 
(\citealt{ma:1979}; \citealt{hu:1999}; \citealt{ib:2001}; \citealt{inan:1999}). 
The high flux, short burst durations and lack of evidence for either 
a binary companion or an accretion disk inspired appeal to unusual energy sources for 
SGRs, including objects with ultra-strong magnetic fields (\citealt{du:1992}), accretion onto 
neutron stars from a fossil disk (\citealt{mar:2001}), and emission from strange quark stars 
(\citealt{bz:2000}).

SGR 1806-20 underwent two major bursting 
episodes since its discovery in 1979, the latest of which occurred in 1996, when {\em Rossi X-ray 
Timing Explorer (RXTE)} observations revealed a 7.47 s pulsar associated with the source that spins 
down at a relatively fast rate of $\sim10^{-11}$ s/s (\citealt{ko:1998}). The detection of pulsations 
and high spin-down rates in the quiescent emission of SGR 1806-20 and SGR 1900+14 (\citealt{ko:1998}; 
1999), when interpreted within the magnetic dipole model, gave strong indication that these 
sources are endowed with ultra-strong magnetic fields ($B>10^{14}$ G) for which the name magnetars was 
coined. However, the lack of independent spectral confirmation of the magnetic field 
strength and the irregularities in the spin-down rates allowed for alternative models 
(\citealt{mar:1999}).

Recently, a 5 keV absorption feature was detected in a burst from SGR~1806-20, providing evidence for 
proton cyclotron resonance in an ultra-strong magnetic field ($B\approx 10^{15}$ G) 
(\citealt{ib:2002}). 
Here we present a comprehensive study of a large number of bursts from the source observed with 
{\em RXTE} during its last active phase in November 1996. We detect the 5 keV feature with high 
confidence in the overall data set. In light of the feature's characteristics and the source's 
observational properties, proton cyclotron resonance is the plausible interpretation, providing the 
first confirmed evidence of proton/ion spectral signature from a cosmic source. We use the feature together 
with the spin-down rate to further constrain the magnetic field and to obtain estimates on the 
gravitational redshift, mass and radius of the SGR.

\section{Observation and Data Analysis}

We obtained the {\em RXTE} Proportional Counter Array (PCA) data from the HEASARC archives 
and used the event-mode and good-xenon data to construct the burst light curves and 
spectra. We detected over 100 bursts with a variety of strengths that range from faint 
bursts, which do not have enough counts to accumulate a meaningful spectrum, to very 
bright bursts, which have large dead-time and pile-up effects and in some cases 
saturated the PCA. For cogent spectral analysis, we considered moderately bright bursts 
which gave enough counts for accurate spectral fitting but at the same time have count 
rate not exceeding 90,000 counts/s so that dead-time and pile-up effects would not be 
significant (\citealt{ib:2001}). Although such instrumental effects are not known to produce 
artificial spectral features but would rather harden the overall spectral continuum and modify the 
widths and depths of real features, this careful selection ensures results that 
are intrinsic to the source.

We classified the bursts according to their temporal profiles into single-peak and multi-peak
bursts and fitted each emission peak separately.
This resulted in a total of 56 spectra\footnote{All burst observations have the same pointing 
coordinates and PCA EDS 
configuration and were analyzed identically.}. 
For background subtraction, we used 
quiescent pre-burst data segments. We found qualitative characterization of the spectral 
continua with an absorbed power-law model, which provided slightly better fits than an 
absorbed bremsstrahlung model for the overall data set. The fits showed a number of 
spectra with relatively large $\chi^2$ and narrow absorption dips in the vicinity of 5 keV. To assess 
the significance of the dip, we added Gaussian absorption lines to the fit 
models and evaluated the change in $\chi^2$ using the F-test. The fits improved considerably, 
giving rise to a highly significant ($\gtrsim\,4\,\sigma$) 5 keV absorption 
feature in some bursts while others showed only modest improvement ($\lesssim\,3\,\sigma$).
Replacing the Gaussian line with an absorption edge was less significant by an order of magnitude.

Fig.~1 shows examples of the bursts with strong evidence of the feature. The spectral 
characteristics of the burst continua and the line features are shown in 
Table 1. All features are narrow; 90\% confidence upper limit on
the line widths range from 0.45 keV to 0.95 keV. The features are fairly 
strong with equivalent widths (EW) varying from 400 eV to 850 eV.

In Fig.~2 we show one of the bursts that did not show the feature significantly, for which we place a 
90\% confidence upper limit on the equivalent width of 200 eV. The fact that these bursts have similar 
temporal profiles and count rates
comparable to those in Fig.~1, yet show no features argues against 
instrumental effects. In fact, the equivalent widths of the features shown in Fig.~1 
exceed by an order of magnitude those of possible artificial features that could be due to 
imperfections in the detector response matrix. As an additional verification, we analyzed 
Crab pulsar spectra of similar counts using an observation that took place just a few hours before those of 
SGR~1806-20. The spectrum is well represented by an absorbed power-law and the fit is not 
improved by adding an absorption line near 5 keV. The 90\% upper limit on the 
equivalent width is 40 eV. 

No evidence is found for a correlation between the appearance of the feature and spectral hardness, 
burst duration, counts, or peak count rate. Interestingly, most of the spectra that showed the feature 
significantly were associated with average 2--50 keV flux in a rather narrow range of $(6.34-9.66)\times10^{-8}$ 
erg cm$^{-2}$ s$^{-1}$ (the flux varied among all spectra between $(4-200)\times10^{-9}$ erg cm$^{-2}$ s$^{-1}$).

To estimate the overall significance of the 5 keV feature in the entire data set, we 
individually fitted each burst spectrum with an absorbed power-law and obtained 
the cumulative $\chi^2_1$ and degrees of freedom $\nu_1$. Next, we added a narrow Gaussian line 
whose energy is allowed to vary around 5 keV only within 1 keV, as implied by the 90\% 
confidence level on the features in Table 1, and fitted the data again and obtained the new 
cumulative $\chi^2_2$ and $\nu_2$. We then performed the F-test on the two $\chi^2$ distributions. 
With ($\chi^2_{1}=2630.3$, $\nu_{1}=2176$) and ($\chi^2_{2}=2302.3$, $\nu_{2}=2063$), we get an overall
chance probability $p=2\times10^{-16}$. If we exclude the five bursts shown in Fig. 1, we still detect the 
feature significantly with a chance probability of $2\times10^{-5}$.

Alternatively, if we assess the overall significance of the feature from the 5 bursts in Fig. 1 using
\footnote{$N$ is the total number of spectra inspected (56), $n=5$,                         
$p_1 ... p_5$ are the chance probabilities given in Table 1, and $p_{min.}$                     
is the least significant $p$.} 
$p={N! \over n! \,(N-n)!} p_1\,p_2\,p_3\,...\,p_n\,(1-p_{min})^{(N-n)}$, we get $p=5\times10^{-14}$.
This is a conservative estimate, not accounting for the presence of the feature in other bursts.

\section{Discussion}

With the growing evidence of ultra-strong magnetic fields in SGRs (\citealt{ko:1998}; 1999) and AXPs
(\citealt{hull:2000}; \citealt{ker:2002}), several models have recently been developed for proton and 
ion cyclotron features (\citealt{bez:1996}; \citealt{za:2001}; \citealt{ho:2001}; \citealt{oz:2002}; 
\citealt{tho:2000}). Recent high-resolution observations of SGRs and AXPs quiescent emission 
revealed no evidence for spectral features (\citealt{ko:2001}; \citealt{fox:2001}; 
\citealt{pat:2001}; \citealt{jue:2002}; \citealt{tie:2002}).
However, two spectral lines at 6.4 keV and 5.0 keV have recently been detected in two 
bursts from SGR~1900+14 and SGR~1806-20, respectively, suggesting that spectral 
features could be a property of the bursts alone (\citealt{st:2000}; 
\citealt{ib:2002}). While the 6.4 keV feature did not provide conclusive evidence on the magnetic 
field strength of SGR~1900+14, the 5.0 keV feature gave evidence for proton cyclotron resonance 
with an implied magnetic field of $\approx\,10^{15}$ G for SGR~1806-20. Our new results reported 
here confirm the 5 keV feature in SGR~1806-20 burst spectra.

The possibility that the 5 keV feature is due to electron cyclotron resonance in a typical 
neutron star magnetic field ($B\approx 10^{12}$ G) is discussed in \citealt{ib:2002}. That the line 
width upper limits are up to an order of magnitude smaller than both observed and expected widths 
of electron cyclotron features argues against this interpretation. Moreover, an electron resonance 
at 5 keV would give a field of $\sim 6\times10^{11}$ G, three orders of magnitude less than that 
derived from the spin-down rate estimate ($\sim 8 \times 10^{14}$~G). The possibility 
that the line could be formed at several neutron star radii from the surface is unlikely due to the
presence of a range of field strengths that would not generate a narrow 
feature. The splitting of Landau levels in $B\sim6\times10^{11}$ G \citep*{pav:1991}, which seems potentially 
promising to distinguish an electron feature from an ion signature if observed with a high 
resolution spectrometer, might be overwhelmed by Doppler broadening. 
Actual measurement of a narrow width would argue for the ion cyclotron interpretation. 

On the other hand, the observed line width, energy, and equivalent width are consistent 
with model predictions for proton cyclotron resonance proposed by \citealt{za:2001} and 
\citealt{ho:2001} for fully ionized Hydrogen atmospheres. 
The magnetic field implied from the proton 
interpretation ($\sim1\times10^{15}$ G) is in good agreement with the independent estimate 
($\sim 8 \times 10^{14}$~G) obtained from the spin-down rate during the same observations 
(\citealt{ko:1998}). 
Lai \& Ho (2002) and \"Ozel (2002) 
pointed out that vacuum polarization effects would reduce the equivalent width of the proton feature 
in quiescent emission. Bulik and Miller (1997) studied the effect of vacuum polarization on the burst 
spectra, but not in connection with the proton resonance. Appropriate models for burst spectra that 
include both effects have not been worked out as yet. 
In conclusion, on observational grounds, a cyclotron resonance at 5 keV together 
with the observed period and spin-down rate of the neutron-star source are consistent only if the 
star's magnetic field is greater than $10^{14}$~G.

In addition to the proton resonance at $E_p$ in $B>10^{14}$~G, $\alpha$--particles and ions 
with $Z/A=1/2$ would resonate in the same field at $E_p/2$. However, the significant increase in the 
binding energy of atoms in a strong magnetic field 
(${\cal E}_o(Z)\propto Z^3~[{\rm ln}\,(B/Z^3)+{\rm const.}]^2$, \citealt{dun:1999}; \citealt{lai:2001}) 
would likely suppress the ion features. 
If the 5 keV feature arises from ion and $\alpha$--particle resonance in $B\sim2\times10^{15}$ G, 
one would expect a {\it similarly strong} feature at $\sim10$ keV due to proton fundamental and 
$\alpha$--particle/ion first harmonics. This is not seen.
There is a hint of a feature near 10 keV in some bursts, but it is more                        
consistent with being a 5 keV second harmonic. Evidence of higher harmonics was seen in the first burst that showed 
the feature (\citealt{ib:2002}).


\begin{deluxetable}{cccccccccc}
\tablecaption{Spectral Characteristics of some SGR~1806--20 Bursts with Strong Evidence For the 5 keV
Absorption Features}
\tabletypesize{\scriptsize}
\tablewidth{0pt}
\tablehead{
\colhead{Burst} & \colhead{Fluence} & \colhead{Photon Index} & \colhead{$N_H$} & \colhead{$\chi_1^2/\nu_1$} 
&
\colhead{E} & \colhead{EW} & \colhead{Width} & \colhead{$\chi^2_2/\nu_2$} &  \colhead{$P_{F-test}$} \\
\colhead{} & \colhead{(counts)} & \colhead{} & \colhead{$(10^{22}\,{\rm cm}^{-2})$} & \colhead{} & 
\colhead{(keV)} &
\colhead{(eV)} & \colhead{(kev)} & \colhead{}
}

\startdata

Burst 1 & 301 & $1.43\pm0.25$ & $20.0\pm7.95$ & 48.2/32 & $5.35_{-0.15}^{+0.21}$ & $850\pm40$ & $<0.47$
& 25.3/30 & $6.3\times10^{-5}$ \\
\tableline

Burst 2 & 1422 & $2.11\pm0.1$ & $22.7\pm2.53$ & 80.2/43 & $5.29_{-0.15}^{+0.16}$ & $560\pm110$ &
$<0.75$ & 44.3/41 & $5.2\times10^{-6}$ \\
\tableline

Burst 3 & 1891 & $1.68\pm0.08$ & $20.3\pm2.19 $ & 58.1/43 & $5.69_{-0.20}^{+0.21} $ & $430\pm100$ & $<0.8$ & 
37.7/41 & $1.4\times10^{-4}$\\
\tableline

Burst 4 & 625 & $1.47\pm0.14$ & $18.5\pm4.12$ & 57.6/43 & $5.29_{-0.20}^{+0.21}$ & $650\pm170$ & $<0.45$ & 
37.4/41 & $1.4\times10^{-4}$\\
\tableline

Burst 5 & 1914 & $1.77\pm0.08$& $20.4\pm2.15$ & 66.5/43  & $5.16_{-0.28}^{+0.28}$ & $400\pm100$ & $<0.95$ & 
49.6/41 & $2.5\times10^{-3}$\\

\enddata
\vspace{-.2 in}

\tablecomments{Errors quoted are the 1$\sigma$ error, except for the energy where the 90\% confidence level 
is given.
The width values are the 90\% confidence upper limit.
Burst 1 is the shortest burst detected with the smallest number of background subtracted counts. With the
C-statistic (Cash 1979), the random probability is $2.7\times10^{-3}$.}
\label{tab:position}
\end{deluxetable}


\begin{figure*} 
\vspace{-.4in} 
\epsscale{0.48}
\plotone{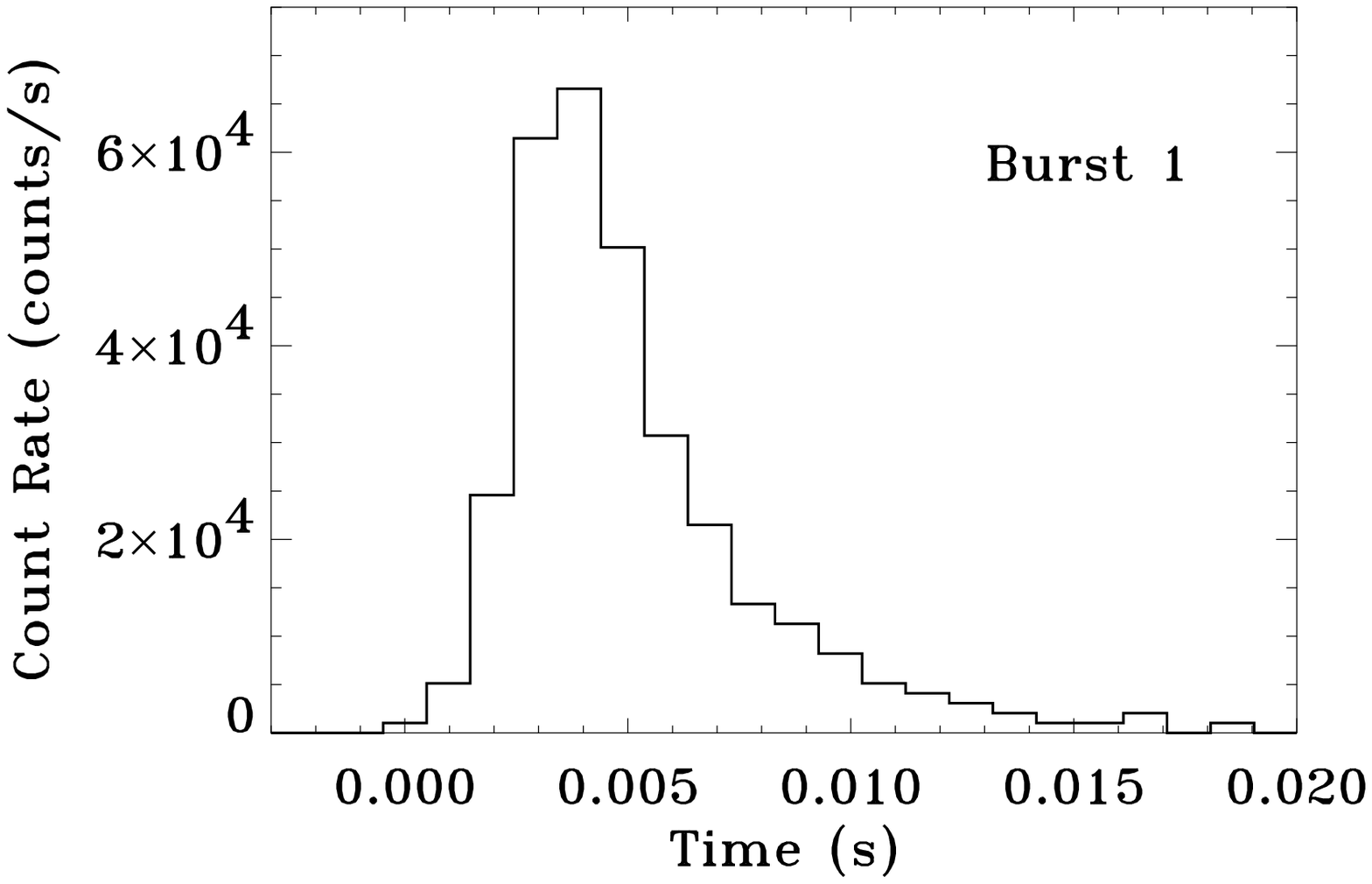}
\epsscale{0.43} 
\plotone{f2.ps} 
\plotone{f3.ps} 
\plotone{f4.ps}
\end{figure*}

\begin{figure*} 
\vspace{-.35in}
\epsscale{0.48} 
\plotone{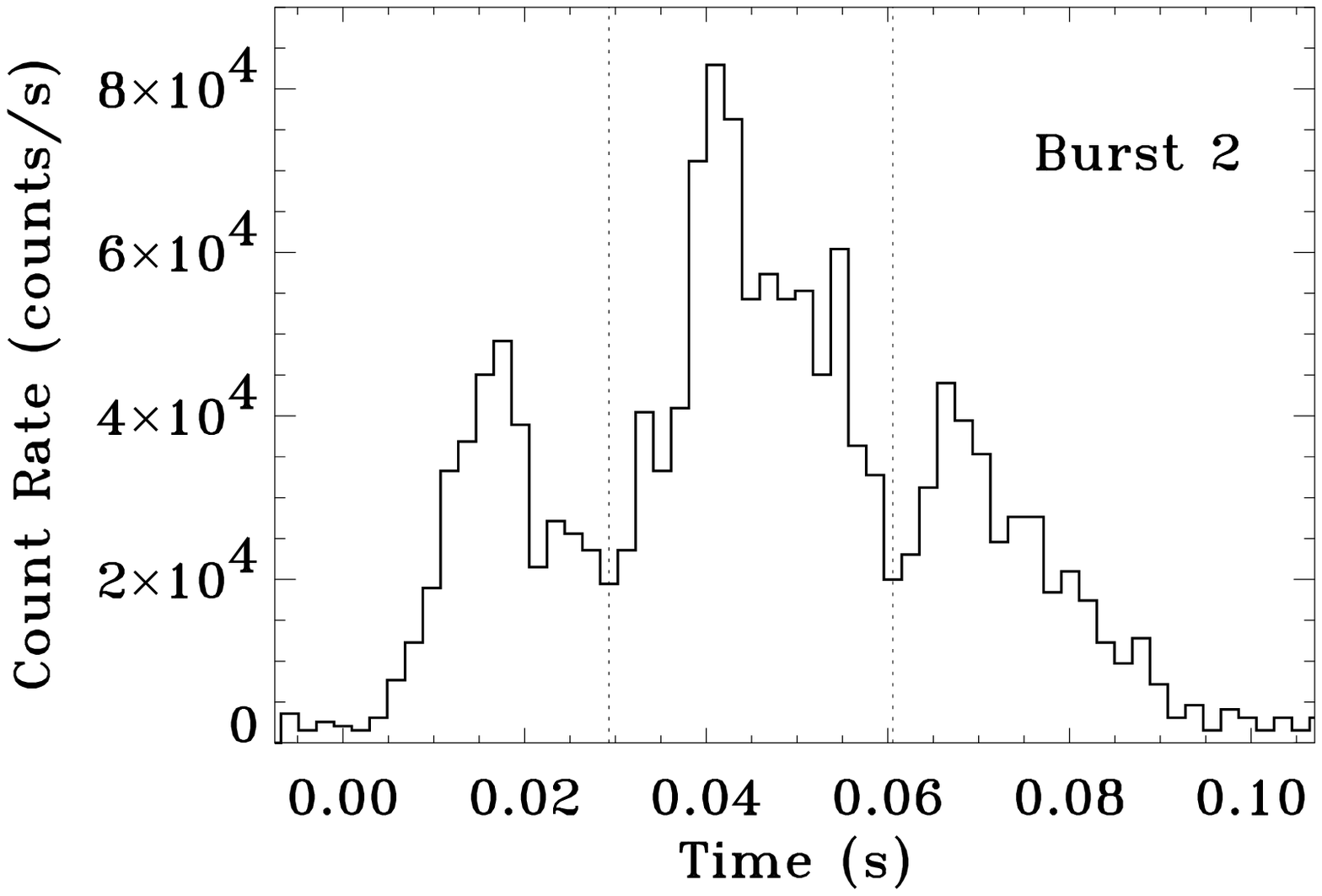}
\epsscale{0.43}
\plotone{f6.ps}
\plotone{f7.ps}
\plotone{f8.ps}
\end{figure*}  

\begin{figure*}
\vspace{-.35 in}
\epsscale{0.48}
\plotone{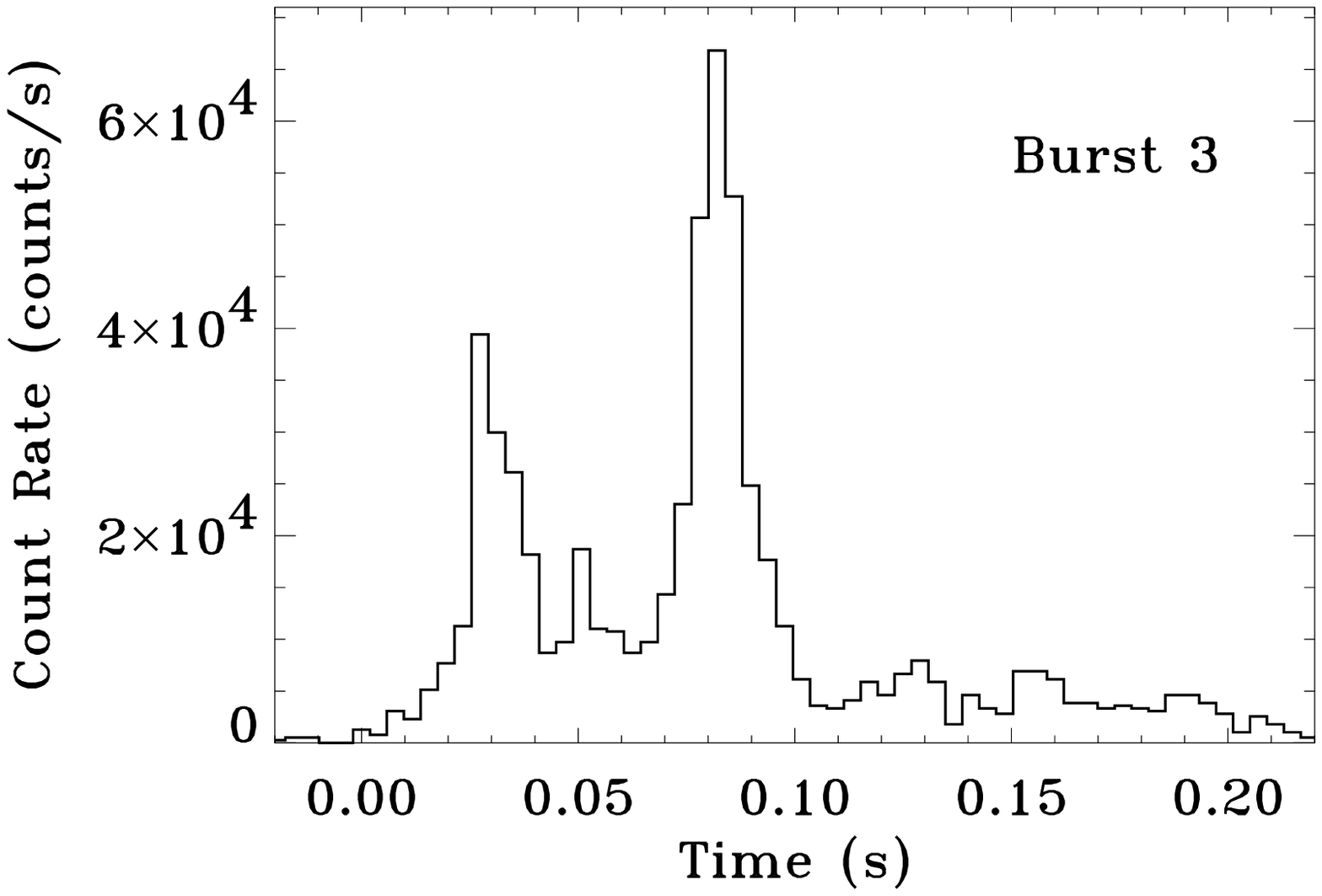}
\epsscale{0.43}
\plotone{f10.ps} 
\plotone{f11.ps}
\plotone{f12.ps}
\end{figure*}

\begin{figure*}
\vspace{-.35 in}
\epsscale{0.48}
\plotone{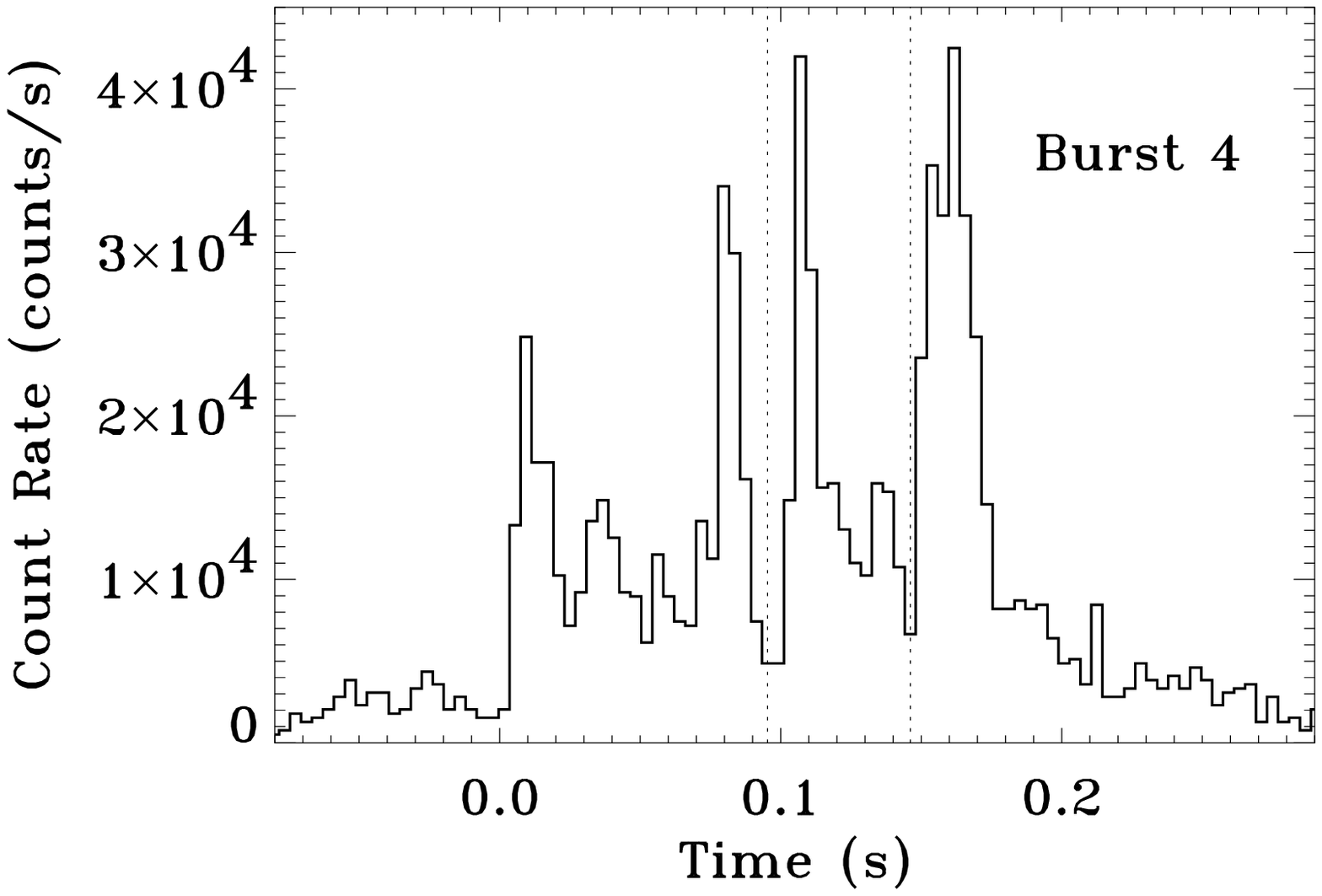}
\epsscale{0.43}
\plotone{f14.ps}
\plotone{f15.ps}
\plotone{f16.ps}
\end{figure*}

\begin{figure*}
\vspace{-.35 in}
\epsscale{0.48}
\plotone{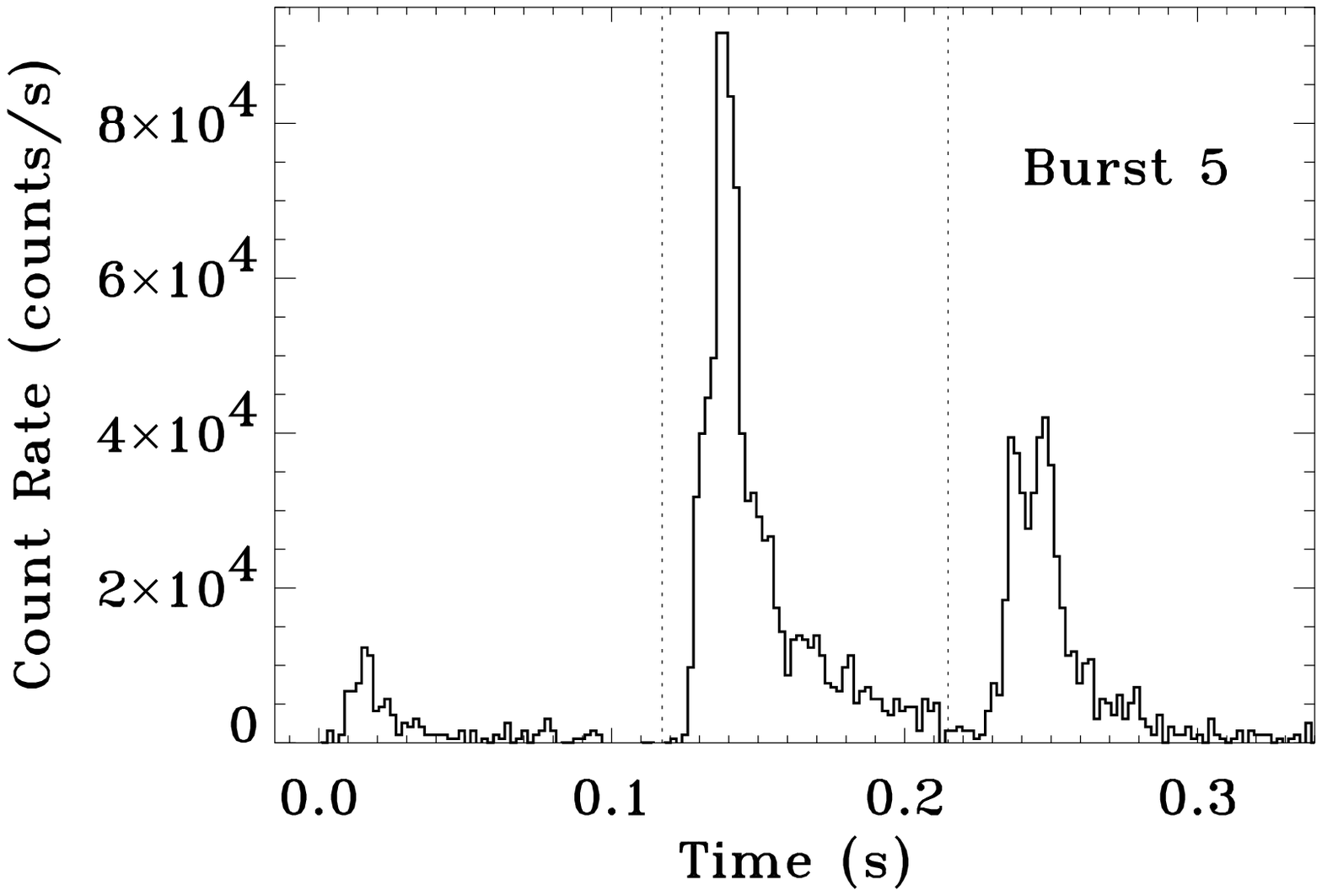}
\epsscale{0.43}
\plotone{f18.ps}
\plotone{f19.ps}
\plotone{f20.ps}

\caption{Bursts from SGR~1806-20 with strong evidence for the 5 keV spectral feature.
The first column panels show the burst light curve profile. The second show the fit residuals with an 
absorbed
power-law model. The third and fourth show the spectrum and residuals, respectively, of the best-fit model
including a narrow Gaussian absorption line. The residuals are shown in units of $\sigma$.
The 5 keV absorption feature is present in the whole of bursts 1 and 3 and in part of bursts 2, 4, and
5. The bursts times are 50405.3704, 50405.5051, 50405.3334, 50405.1843, 50405.4675 (MJD), respectively.
}
\end{figure*}


\begin{figure*} 
\epsscale{0.48}
\vspace{-.28in} 
\plotone{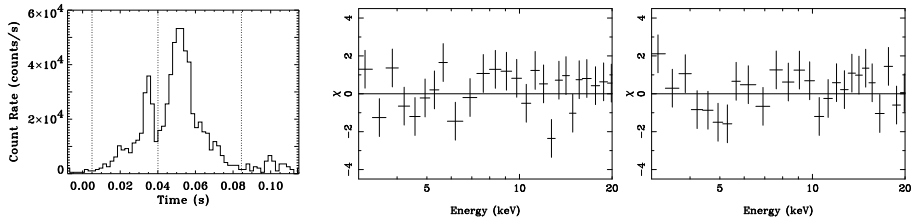}
\epsscale{0.43} 
\plotone{f25.ps}
\plotone{f26.ps}
\figurenum{2}  
\caption{An example burst with modest or no evidence for the 5 keV feature.
The two-peak burst shows random residuals in the first peak and weak absorption excess
($2.4\,\sigma$) in the second peak.  The burst time is 50405.5191 (MJD).
}
\end{figure*}

Using the proton fundamental resonance formula $E_p = 6.3\,(1+z)^{-1} (B/10^{15}\,{\rm 
G})$~keV
\footnote{Here $z$ is the gravitational redshift and $(1+z)^{-1} = (1-2GM/Rc^2)^{1/2}$, 
and $M$ and $R$ are the mass and radius of the star, respectively.}, 
and the observed resonance values, the magnetic field in SGR~1806-20 is directly determined by $B = 
(7.7-9.4)\,(1+z)\times10^{14}$~G.
For a given $E_p$ and $B$, the resonance relation gives a locus of $M$ and $R$. 
Similarly, the magnetic field--spin-down relation
\footnote{$P$ and $\dot{P}$ are the period and spin-down rate, $\theta$ is the angle between 
magnetic dipole and rotation axis, and $\xi$ is the neutron star's moment of inertia in units of 
$MR^2$.}, 
$B^{2}={3\,c^3\,\xi\,P\,\dot{P}\over 2\,\pi^2\,{\rm sin}^2\theta}\,{M\over R^4}$, gives another 
locus of $M$ and $R$ for a particular $B$. The intersection of the two relations constrains the 
mass and radius of the SGR as illustrated in Fig.~3. 
While the dipole model comes from an idealized case, it provides adequate order of 
magnitude estimate for the magnetic field. Also, since the connection of SGR~1806-20 and SNR G10.0-0.3 
is now uncertain (\citealt{ga:2001}), there is no longer a requirement that it has a large wind of 
particles to fill a plerionic nebula. It is then possible for the magnetic dipole radiation to be the 
dominant source of spin-down.


\begin{center}  
\includegraphics[scale=0.45]{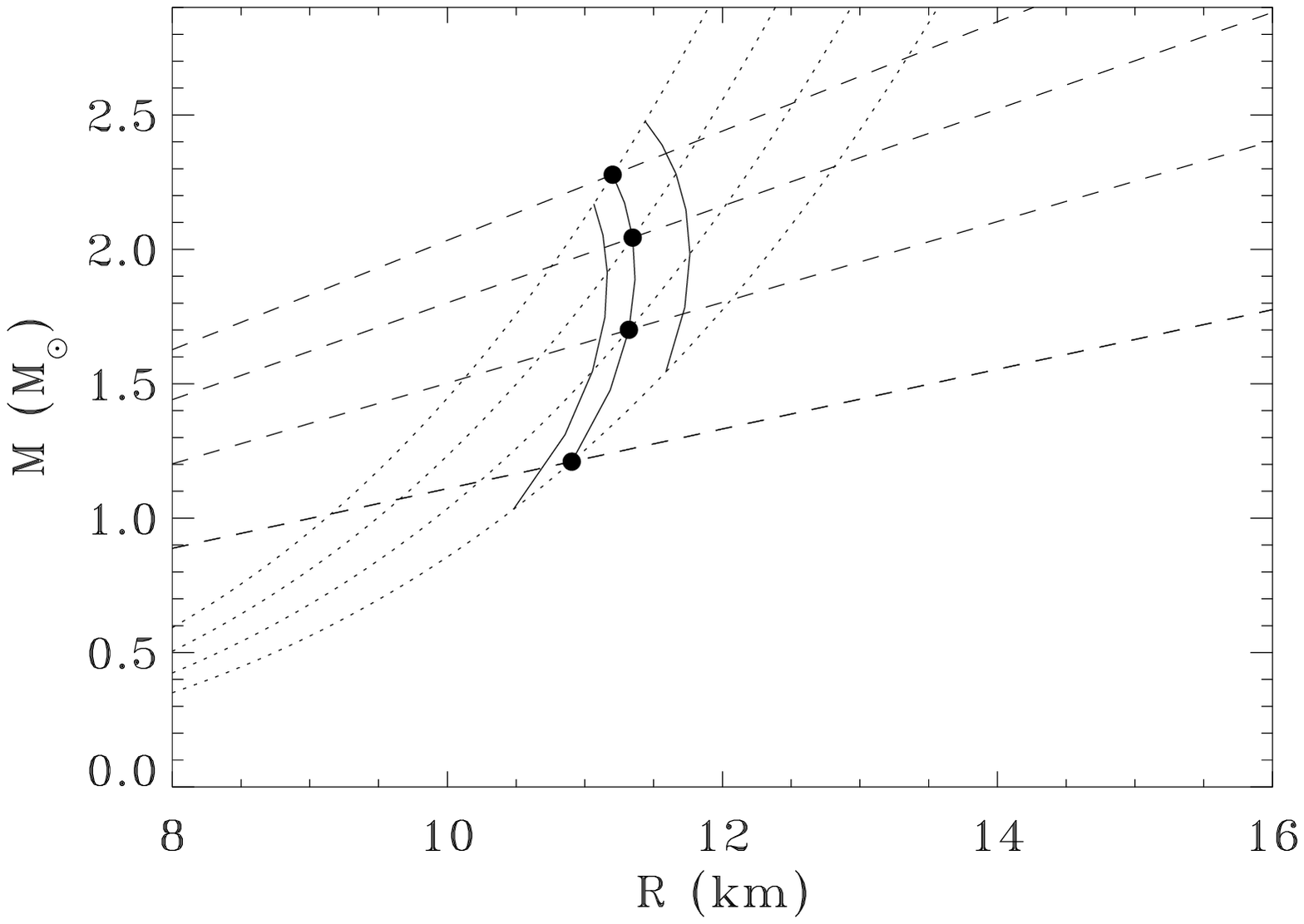}
\figurenum{3}   
\figcaption{Mass-Radius curve for SGR~1806-20 as obtained from the intersection of the proton cyclotron 
resonance
relation (dashed lines) and the spin-down rate relation (dotted curves) for an oblique rotor model. Both 
relations
are plotted at $B=(1.0-1.3)\times10^{15}$ G (increasing from bottom to top). The line along the
intersection points corresponds to the best-fit cyclotron resonance energy while the other two refer to
the 90\% confidence level.
Here we used the feature with the largest redshift ($E_p=5.16$ keV) and assume that it comes from very close 
to the
stellar surface. For equations of state of current interest, $\xi$ would be in the range 0.3-0.45 for
$0.1 < (M/M_\odot)/(R/{\rm km}) < 0.23$ (\citealt{latt:2001}).}
\end{center}

Assuming an oblique rotator and the average value (0.5) of ${\rm sin}^2\theta$,
the $M$-$R$ constraints above, along with the equations of state, would limit the radius and 
mass of SGR~1806-20 to $R=(10.5-12.3)$~km and $M=(1.3-1.8)\,M_\odot$ if the neutron star were 
composed of nucleons and hyperons. The corresponding gravitational redshift and mass-radius
ratio are $z=0.26-0.33$ and $M/R=(0.12-0.15)\,M_\odot$/km, which yield a surface 
magnetic field $B=(0.97-1.3)\times10^{15}$ G. We note that our estimates of $z$ and $M/R$ are 
close to those recently reported from an isolated neutron star that showed 0.7 and 1.4 keV 
absorption features (\citealt{san:2002}). For ${\rm sin}^2\theta=1$, the $M$-$R$ 
locus moves to values of $R<10$ km, too small for equations of state that do not 
have quarks or other `exotic' particles. A quark star would give $R=(8-10)\,$km and $M=(1.3-1.95)\,M_\odot$.

With our present knowledge, there are several mechanisms that could explain the observed features
in light of the magnetar model (\citealt{tho:1995}). SGR bursts have been identified with
photon-pair plasma that is created and trapped in the magnetosphere as a result of magnetically
driven crust-quake. The plasma dissipates its energy by X-ray emission; part is radiated away as
the burst and part is radiated back to the exposed surface of the neutron star. In this framework,
proton cyclotron resonance could arise in several scenarios.

First, it could originate near the surface of the neutron star, in the local region of the 
crust-quake, where hydrogen in the atmosphere is being irradiated by the plasma photons. 
Secondly, the cooling phase of the burst emission is expected to involve an ablation process 
that blows off baryons and ions from the surface of the star to its atmosphere. If 
this occurs in the first peak of a multi-peak burst (or in a single-peak burst), the optical depth 
during a subsequent peak (or a later burst) is likely to be larger, thus augmenting 
resonance features.

Features produced by these mechanisms come from near the surface and thus 
can be obscured by the rotational orientation of the star if the emission region is not along our line of sight.
Note that the durations of the absorption line episodes are much smaller 
($<0.01~P$) than the pulsar rotational period. It was recently pointed out that emission 
from near the equatorial plane will be pronounced at the cyclotron resonance, in contrast with that 
from the poles (\citealt{tho:2002b}).

A third origin of proton resonance is from the baryon-loaded sheath, 
which constitutes the plasma outer shell. The resonance could occur in that region as 
protons in the baryon sheath are illuminated from within by the hard burst photons.

In all the scenarios above the burst is essential, giving rise to optical depths and fluxes of high 
energy photons far exceeding those in quiescence, which may explain the absence of the feature from 
SGR and AXP persistent emissions. The effect of vacuum polarization, together with the correlation 
with the flux, the rotational and latitude constraints mentioned above, and the fact that the 
optical depth to resonant cyclotron scattering is proportional to the local density of protons 
could in principle account for the transient nature of the feature and its absence or weak presence 
in some bursts.

In summary, the discovery of the 5 keV absorption feature provides a unique opportunity 
not only to diagnose SGR~1806-20 magnetic field but also to reveal insight on the 
gravitational field strength, mass and radius of SGRs. 
This is an important step toward understanding a mysterious constituent of the Universe.
Further observations and careful analysis could allow unprecedented studies of physical phenomena 
in the extreme limits of gravitation and magnetic field, uniquely found in magnetars.

We are grateful to R. Duncan, C. Markwardt, K. Hurley, S. Safi-Harb, S. Zane, R. Turolla, F. 
\"Ozel, G. Pavlov, C. Miller, K. Jahoda and A. Harding for useful discussions and insightful 
comments. A.I.I. acknowledges support from NASA grant No. NAG5-11951 and The George                             
Washington University. 


\end{document}